# Positive Ion Impediment by short-circuiting effect in a magnetized plasma column

Satadal Das[1,2,a)] and Shantanu K. Karkari[1,2,b)]

[1]Institute for Plasma Research, Bhat, Gandhinagar, Gujarat 382428, India
[2]Homi Bhabha National Institute, Training School Complex, Anushakti Nagar, Mumbai 40094, India

E-mail: [a)]satadal.das@ipr.res.in, [b)]skarkari@ipr.res.in

## Abstract

The radial characteristic of a partially magnetized plasma column created by a hot cathode filament is presented. It is found that in the absence of magnetic field, plasma potential and density varies similarly according to Boltzmann distribution. However when magnetic field increases, a clear divergence is seen as the plasma density becomes more pronounced in the centre whereas a corresponding minima is observed in plasma potential; which impede the radial diffusion of positive ions towards the grounded sidewalls. A phenomenological model based on short-circuiting effect is developed, which fairly explains this contrasting behaviour.

1. ## Introduction

The factors affecting the spatial density and potential profile inside a magnetized plasma system is an open problem in plasma physics [1 - 10]. Over the past 70 years, this topic has remained at a centre stage due to its significance in fusion research as well as in industrial applications of plasma. The magnetic field introduce remarkable effect on spatial density or potential characteristics. This is caused due to significant disparity in the transport rate of electrons and positive ions across the magnetic field. The intrinsic electric fields generate $\vec{E} \times \vec{B}$ drifts, which lead to cross-field turbulent transport [10, 11, 12]. The validation of Bohm criteria [13] and assumption of Non-Boltzmann distribution for electrons across magnetic field have stimulated by researchers to deeply investigate the behaviour of magnetized plasma [7, 14, 15].

In laboratory plasma setup, the magnetic field usually intercept the grounded experimental chamber or other metallic electrodes. This leads to a preferential loss of plasma electrons along magnetic field lines to the conducting walls. The difference in cross-field mobility of electrons and positive ions lead to a non-ambipolar transport across magnetic field. In 1955 Simon [16] proposed that anomalously large diffusion rate observed in magnetized plasma systems is rather due to short-circuiting effect by the conducting end-plates, which ubiquitously exists in such bounded plasma system. He further emphasized that exotic assumption such as Bohm diffusion [13], which yields, $D_\perp \sim B^{-1}$ as compared to classical diffusion $D_\perp \sim B^{-2}$ need not be necessary to explain the unusual large diffusion rate presumably caused by $\vec{E} \times \vec{B}$ induced charge particle fluctuations across magnetic field [16, 17, 18]. Nevertheless the Bohm diffusion has been validated in numerous cases [19, 20].

The short-circuiting phenomena proposed by Simon have been recently recalled in a few publications [19, 21, 22, 23]. In ref. 24, Chen and Curreli have invoked this concept to assume electrons to be obeying Boltzmann distribution across the magnetic field and concluded that the plasma density profile in a cylindrical system is always peaked at the centre. They argued that for a conducting wall, an escaping ion across the magnetic field is neutralized by absorbing an electron to the wall; whereas the parallel flow of electrons along the magnetic field is restricted by developing a negative potential giving rise to a retarding sheath; which helps in maintaining an ambipolar flow across the magnetic field. The position of the primary ionization source inside magnetized plasmas is also found to influence both radial density and plasma potential in magnetized plasma devices [25, 26, 27]; however adequate discussion on the role of experimental system on the equilibrium plasma properties is rather seems to be limited.

The significance of the above problems has motivated to develop various theoretical models to describe the radial density and potential profile inside magnetized plasma column [14, 15, 28]. The radial plasma potential and density profile in such systems is governed by the diffusion equations, with coefficients which are dependent on collisions between the charge species or neutrals as well as magnetic field strength. In ref. 28, different diffusion cases has been considered; wherein the non-ambipolar flow across the magnetic field represents the short-circuit limit described by Simon [16]. This condition requires that the plasma length should be larger than the plasma radius so that the positive ions move largely in the radial direction, whereas the highly mobile electrons flow axially along the magnetic field [28]. A comprehensive model taking into account the effect of ionization, ion and electron inertia, respective collisions with gas atoms has been demonstrated in Ref. 14. They defined a parameter, $= \frac{R \lambda_e}{\rho_e \rho_i}$ ; which sets the condition whether the Boltzmann distribution can be applied for electrons across magnetic field. The independent models developed in ref. 14 and 28 suggest a Bessel like solution for plasma density, having a peak at the centre whereas the plasma potential tends to increase steeply near the wall. The retarding potential for the ions helps to impede them at the centre.

In-spite of remarkable theoretical works, the laboratory experiments to validate the models is nearly absent. In this paper, we present an experimental study of radial plasma density and potential behaviour inside a magnetized plasma column. The experimental results clearly indicate that the radial density distribution behaves oppositely to the variation in plasma potential inside the cylindrical plasma chamber. Although the radial plasma density is found to be consistent with previous reported results; however the radial plasma potential shows a distinct trend, though it increases radially outward [14, 28]. To understand this effect, a phenomenological model has been proposed, which considers the short-circuiting effect in to account for determining the potential profile.

The paper has been organized as follows. The experimental setup is briefly described in Section-2. In section-3, the phenomenological model has been presented along with the experimental results. The important results are further discussed in Section-4 and the outcome of the work has been summarized in Section-5.

2. **The Experimental Setup:**

The experiment is carried in a cylindrical vacuum chamber shown in Fig. 1(a). The chamber is evacuated to a base pressure $2 \times 10^{-5}$ mbar with 500/s diffusion pump and backed by a rotary pump. The axial magnetic field is produced by a pair of electromagnet coils made from copper cables by winding in double pan-cake configuration. The coils are assembled in Helmholtz configuration to produce axial magnetic field B = 800 Gauss at the centre for coil current 60 A. Fig. 1(b) shows the axial magnetic field plots. The highlighted region shown on the plots corresponds to uniform magnetic field region over a distance of 20 cm.

The plasma is produced in argon with a pair of hot tungsten filament, which acts as cathode. The tungsten filaments (diameter 0.25 mm and length 10 cm) are heated by passing 10 A – 12 A alternating current at 50 Hz, provided by a step-down transformer. For extracting the electrons from the filament, the centre tap of transformer is biased at -150 V with respect to the grounded chamber. In the entire experiment, the pressure was kept at 1 – 2 Pa.

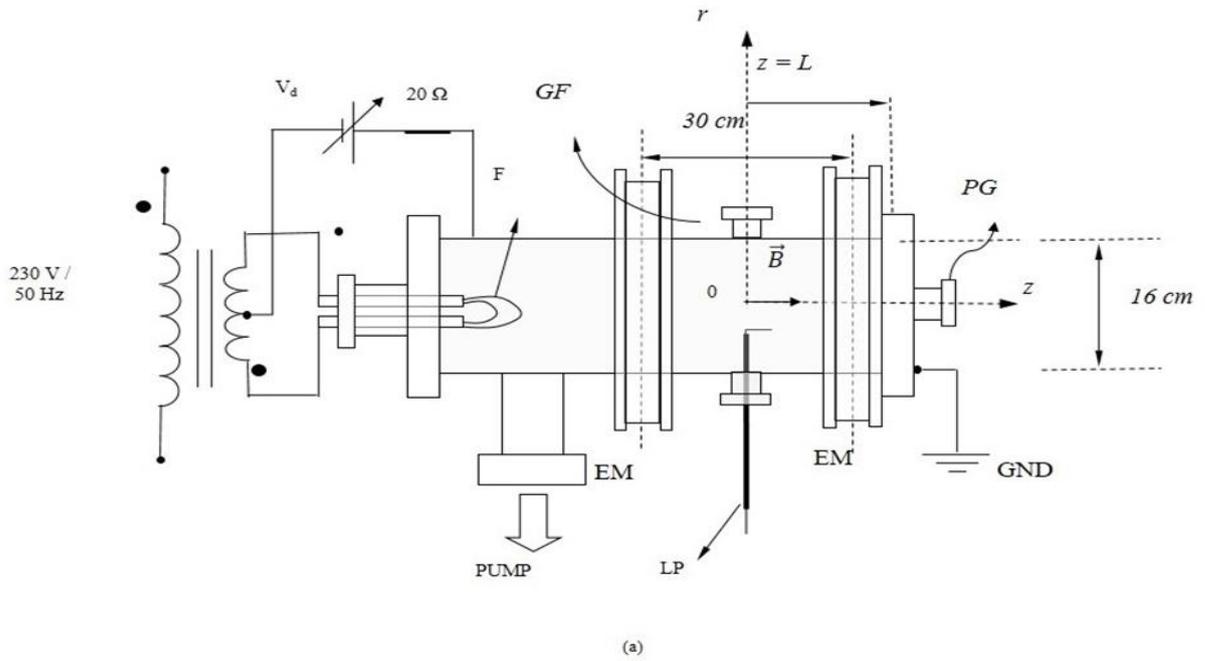

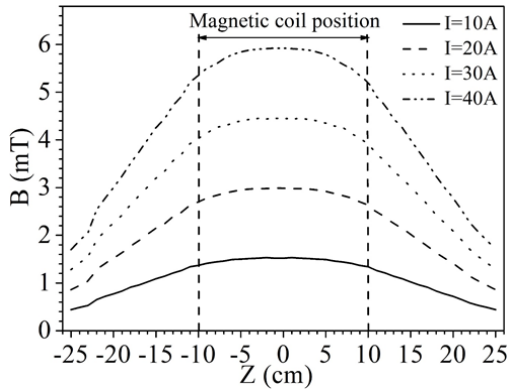

Fig.1(a): Schematic of the experimental setup: EM – Electro-magnets, $V_d$ – Discharge voltage, F – Tungsten filament, G – Gas feed valve, PG – Pressure gauge, GND – Ground

Fig.1(b): Plot of axial magnetic field between the electro-magnet coils; shaded region shows the region between the EM coils.

## 3. Phenomenological model and the Experimental results:

In the experiment, the plasma is created by the ionizing electrons produced from the filament, which are accelerated in all direction towards the grounded walls of the chamber. In the absence of magnetic field, the plasma is seen to fill out the entire volume. However when the axial magnetic field is introduced, the plasma in the uniform magnetic field region seems to be brighter near the centre/ along the axis. Ignoring the sheaths at the grounded walls, the overall plasma is largely uniform along the magnetic field. In order to model the plasma in the uniform magnetic field region between the electro-magnets, we assume cylindrical plasma tube, with axial magnetic field as shown in figure-2(a).

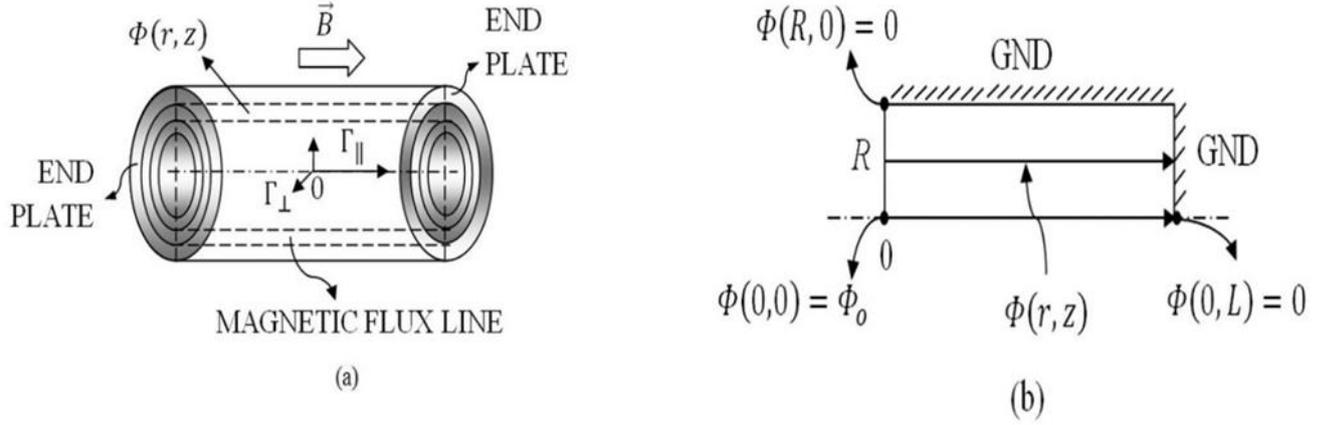

Fig- 2(a) Schematic of magnetized plasma column showing equipotential plasma Ø(r,z) at any point along the magnetic field flux surfaces; (b) figure highlighting a rectangular cross-section of the cylindrical column, with magnetic field along the z-axis.

The cylindrical plasma column is shown to terminate at the conducting end plate at one end as well as radially at the grounded chamber walls. The arrows indicate the direction of the parallel and the perpendicular electron fluxes with respect to the magnetic field. Since $\Gamma_\parallel \gg \Gamma_\perp$, therefore the electrons tends to reach the grounded endplate rapidly. Since the radial walls are also grounded, therefore the potential at any point measured with respect to the grounded end plate gives the measure of radial plasma potential drop in that magnetic flux tube.

To simplify the above problem, we further consider a small cross-section of the cylinder with origin defined at the centre of the plasma column as shown in the figure 2(b). The potential at different regions are indicated in the figure.

a. *Radial Plasma Density:*

The steady state plasma density is calculated by solving the flux continuity equations for positive ions and electrons which are given by;

$$\vec{\nabla} \cdot \vec{\Gamma_i} = S, \qquad \vec{\nabla} \cdot \vec{\Gamma_e} = S \qquad (3.1)$$

Here S is the source term due to ionization and $\vec{\Gamma_i}$ and $\vec{\Gamma_e}$ are the ion and electron flux density. By using the expressions used in Ref.2 for $\Gamma_e$ and $\Gamma_{ion}$, the above equation can be expanded in cylindrical coordinates as written in Eqs. (3.2) and (3.3). In the above equations, the axial density and potential variation have been considered to be small; i.e, $\frac{d^2 n}{dz^2} = 0$. The $\nu_e$ and $\nu_i$ are the ionization frequency for electron and ion respectively,

$$D_{\perp i}\frac{d^2 n_i}{dr^2} + \frac{1}{r}D_{\perp i}\frac{dn_i}{dr} + \frac{D_{\perp i}}{T_i}\frac{d}{dr}\left(n_i \frac{d\emptyset}{dr}\right) + \frac{D_{\parallel i}}{T_i}\frac{d}{dz}\left(n_i \frac{d\emptyset}{dz}\right) = -\nu_i n_i \qquad (3.2)$$

$$D_{\perp e}\frac{d^2 n_e}{dr^2} + \frac{1}{r}D_{\perp e}\frac{dn_e}{dr} - \frac{D_{\perp e}}{T_e}\frac{d}{dr}\left(n_e \frac{d\emptyset}{dr}\right) - \frac{D_{\parallel e}}{T_e}\frac{d}{dz}\left(n_e \frac{d\emptyset}{dz}\right) = -\nu_e n_e \qquad (3.3)$$

For a typical experimental parameters such as $T_e$ =2eV, $T_i$ =0.025eV, $\nu_e \sim 10^3$ Hz, $\nu_i \sim 10$ Hz, $\frac{\nu_e}{\nu_i} \sim 10^2$; the diffusion coefficients, is found to be; $D_\perp \ll D_\parallel$. Assuming the parallel electric field is localized within the sheath and the plasma is quasi-neutral, $n_i \cong n_e = n$ along the magnetic flux lines; hence the axial electric field $\frac{d\emptyset}{dz}$ term can be ignored in Eqs (3.2) and (3.3).Therefore the fourth term in Eqs (3.2) and (3.3) can be neglected [16]. Therefore;

$$D_{\perp i}D_{\perp e}(T_i + T_e)\frac{d^2 n}{dr^2} + D_{\perp i}D_{\perp e}(T_i + T_e)\frac{1}{r}\frac{dn}{dr} = -(\nu_e T_e D_{\perp i} + \nu_i T_i D_{\perp e})n \qquad (3.4)$$

Also from the above experimental parameters, $T_i \ll T_e$, so $T_i$ and $\nu_i T_i D_{\perp e}$ can be neglected in Eq. (3.4). This gives the expression for the radial density variation as follows,

$$\frac{d^2 n}{dr^2} + \frac{1}{r}\frac{dn}{dr} + \frac{\nu_e}{D_{\perp,e}}n = 0 \qquad (3.5)$$

The solution of Eq. (3.5) is expressed as,

$$n(r) = n_0 J_0(\gamma r) \qquad (3.6)$$

Where, $J_0(\gamma r)$ is zeroth-order Bessel's function.

Considering the boundary condition $n(r = R) = 0$, where $R$ is the radius of chamber, we obtain the parameter $\gamma$

$$\gamma = \left(\frac{\nu_e}{D_{\perp,e}}\right)^{\frac{1}{2}} \qquad (3.7)$$

As $\omega_{ce}$ is in GHz and $\nu_c$ in MHz range for the magnetic field strengths, B = 4.58 to 6.53 mTesla, $\omega_{ce} \gg \nu_c$; hence the coefficient $D_{\perp,e} \approx \frac{k_B T_e \nu_c}{m_e \omega_{ce}^2}$ ; where $\nu_c$ and $\omega_{ce}$ are electron neutral collision frequency and electron cyclotron frequency respectively. The normalized plasma density can be written,

$$N(r) = J_0(\gamma r) \qquad (3.8)$$

Eq. (3.8) gives the radial density distribution inside the discharge. The effect of magnetic field is absorbed implicitly in $\gamma$.

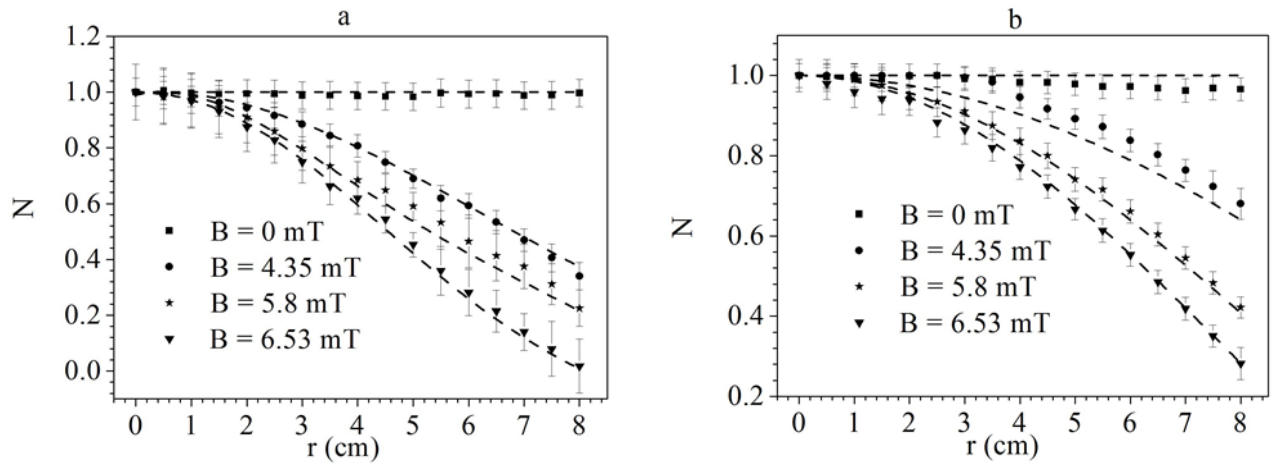

Fig-3: Graph of radial plasma density variation for P = 1.2Pa (a) and 1.7Pa (b). Dash lines are theoretical plots and scattered points are experimental plots.

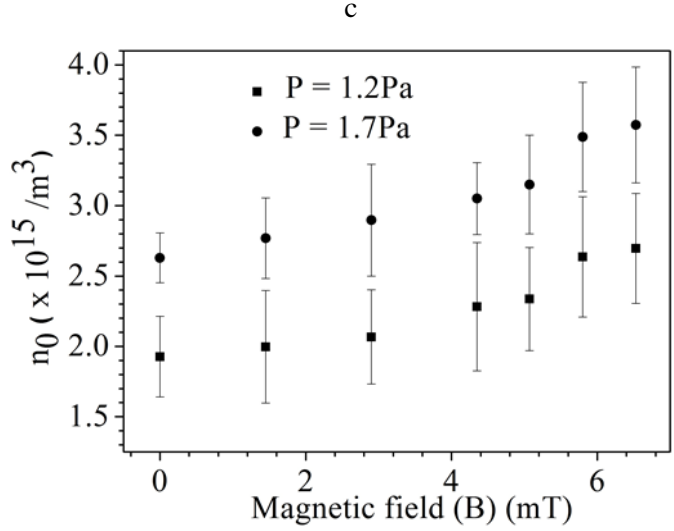

Fig-3.c: Graph of central density variation with axial magnetic field.

The radial density profile with magnetic fields in Fig.3 (a, b) shows that the density gradient increases with the magnetic field strength. This is because of the magnetic field confines the primary ionizing electrons and it enhances the plasma density in the central region. The central plasma density corresponding to the above figures are plotted as a function of the magnetic field as shown in Fig.3(c).

b. *Radial Plasma Potential:*

Considering the plasma electrons as Boltzmann distribution [29], which is expressed as;

$$n_e = n_0 \exp\left[\frac{e(\phi-\phi_0)}{k_B T_e}\right] \tag{3.9}$$

Where, $\phi_0$ is the central potential. So potential variation for un-magnetized plasma can expressed as;

$$\phi(r) = \phi = \phi_0 + T_e \ln\left[\frac{n(r)}{n_0}\right] \tag{3.10}$$

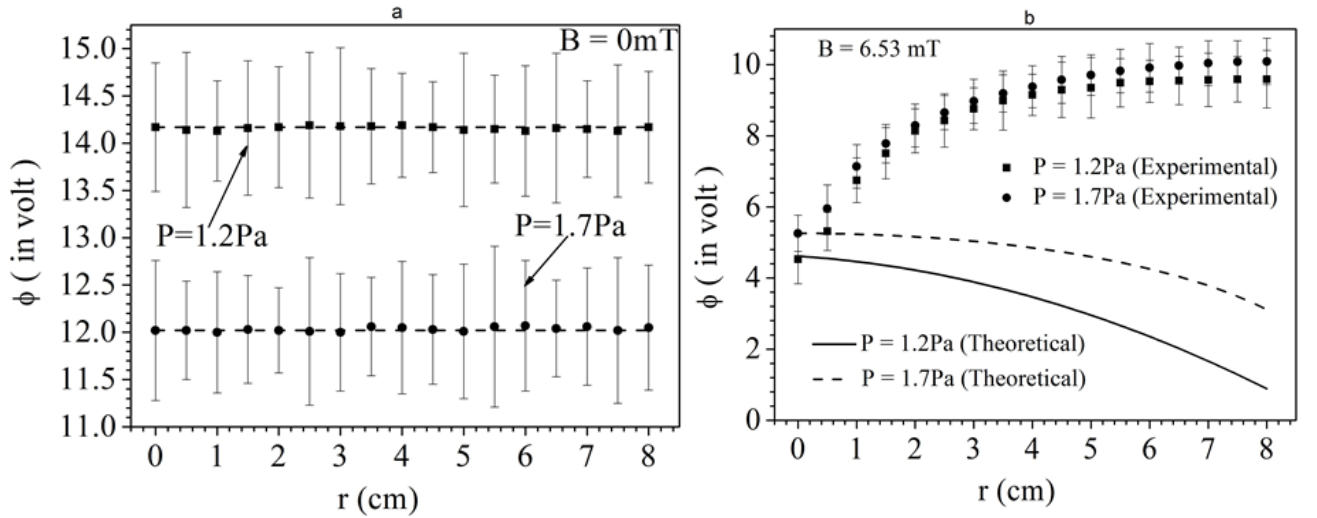

Fig-4: Graph of radial potential variation using Boltzmann Distribution for un-magnetized (a) and for magnetized plasma (b) at P =1.2Pa and 1.7Pa. Dash and Solid lines are theoretical plots and scattered points are experimental plots.

As shown in Fig.4, for un-magnetized case the plasma potential obtained in experiment match perfectly according to Eq. (3.10), however a highly contrasting trend is observed during the presence of magnetic field. This observation confirms that the Boltzmann distribution is not valid across magnetic field, as also reported in Ref. 14, 28. The radial potential profile obtained in the experiment shows a distinct trend than the analytical results in Ref.14, which found a steep rise in $V_P$ towards the wall. In the present case the plasma potential tends to saturate with radial distance from the centre. To address this issue, we revisit the above problem as follows:

Since the electrons can readily flow along the magnetic field, therefore it is reasonable to assume that the Boltzmann distribution is independently valid for electrons along the magnetic field lines. We further consider that in the present setup, the axial magnetic field intercepts the grounded chamber flange (z = L) as shown in Fig. 1(a). Therefore the plasma density distribution at anywhere inside a given magnetic flux tube should be related according to;

$$n_e(r,z) = n(r,0)\exp\left[-\frac{e[\emptyset_0(r)-\emptyset(z)]}{k_B T_e}\right] \qquad (3.11)$$

Where, $\emptyset_0(r)$ refers to the central potential in the individual flux tubes and $\emptyset(z)$ is the local potential along z-axis in that flux tube.

Since the central density $n(r,0)$ in Fig.3 (a, b) is shown to vary as $n(r,0) = n_0 J_0(\gamma r)$, hence Eq. (3.11) can be generalized to,

$$n_e(r,z) = n_0 J_0(\gamma r)\exp\left[-\frac{e[\emptyset_0(r)-\emptyset(z)]}{k_B T_e}\right] \qquad (3.12)$$

In the above equation, all the potentials are measured with respect to the ground. Therefore at the wall, $\emptyset(z=L) = 0$.

Hence Eq. (3.12) can be written as;

$$n_e(r,L) = n_0 J_0(\gamma r)\exp\left[-\frac{e\emptyset_o(r)}{k_B T_e}\right] \qquad (3.13)$$

From the above, the electron flux at the grounded end-plate along the magnetic field can be written as:

$$\Gamma_{\parallel e} \approx \Gamma_{eo} J_0(\gamma r)\exp\left[-\frac{e\emptyset_o(r)}{k_B T_e}\right] \qquad (3.14)$$

Where, $\Gamma_{eo} = n_0 v_{th}$ is the electron flux defined at the centre of the discharge, i.e. (r=0, z=0).

$$v_{th} = \sqrt{\frac{3k_B T_e}{m_e}}$$ is the electron thermal speed.

Considering that the plasma density has a radial fall towards the wall, therefore according to Eq. (3.14) the peripheral region of the end plate will receive lesser electron flux than at the centre. According to Simon short-circuit effect the positive ions can easily move across the magnetic field lines as compared to electrons. Thus the potential difference created between the central region $\emptyset_0(0)$ and the periphery $\emptyset_0(r)$ will eventually drive a current through the grounded end plate as illustrated in Fig. 2(b). Since the short-circuited current is constituted mainly by electrons flowing axially along the magnetic field lines to the end plates, therefore according to Boltzmann distribution, the parallel electron flux can be expressed as follows:

$$\Gamma_{\parallel e} \approx \Gamma_{eo} J_0(\gamma r)\exp\left[-\frac{e\{\emptyset_o(r)-\emptyset_o(r=0)\}}{k_B T_e}\right] \qquad (3.15)$$

Since the radial movement of electrons inside the plasma is highly limited by the axial magnetic field, therefore the flow is non-ambipolar across the magnetic field. However according to Simon short-circuit effect [16] a combined ambiplority can be achieved when the following condition is satisfied;

$$[\Gamma_{\parallel e} + \Gamma_{\parallel i}] = -[\Gamma_{\perp e} + \Gamma_{\perp i}] \tag{3.16}$$

Considering that, $\Gamma_{\parallel e} \gg \Gamma_{\parallel i}$; the above equation reduces to $\Gamma_{\parallel e} \approx -(\Gamma_{\perp e} + \Gamma_{\perp i})$. On substituting the expressions for different fluxes, $\emptyset_0(r)$ as $\emptyset(r)$ and $\emptyset_o(r=0)$ as $\emptyset_o$, we get;

$$n_0 J_0(\gamma r) \exp\left[\frac{e\{\emptyset_0 - \emptyset(r)\}}{k_B T_e}\right] v_{th} = -[-\mu_{\perp e} n E_\perp - D_{\perp e}\nabla_\perp n + \mu_{\perp ion} n E_\perp - D_{\perp ion}\nabla_\perp n] \tag{3.17}$$

Considering the un-magnetized ions are at room temperature, it can be seen that $\mu_{\perp e} \ll \mu_{\perp i} \sim \mu_{\parallel i}$ as well as $D_{\perp e} \ll D_{\perp i} \sim D_{\parallel i}$. This further simplifies Eq. (3.17) which finally reduces to,

$$v_{th}\exp\left[\frac{e\{\emptyset_0-\emptyset(r)\}}{k_B T_e}\right] = \frac{\mu_{\perp i} N}{J_0(\gamma r)}\frac{d\phi(r)}{dr} + \frac{D_{\perp i}}{J_0(\gamma r)}\frac{dN(r)}{dr} \tag{3.18}$$

$$v_{th}\exp\left[\frac{e\{\emptyset_0-\emptyset(r)\}}{k_B T_e}\right] = -\frac{\mu_{\perp i} N \gamma J_1(\gamma r)}{J_0(\gamma r)}\frac{d\emptyset(r)}{dN} \tag{3.19}$$

The first derivative of the zeroth order Bessel function, $\frac{d}{dr}J_0(r) = -J_1(r)$; $J_1(r)$ = First order Bessel function. In view of the above model, the final expression for the radial potential profile can be obtained as follows:

$$\emptyset(r) = \emptyset_0 - T_e\left[\ln\left(\frac{\mu_{\perp i}\gamma}{v_{th}}\right) + \ln(T_e) + \ln\left\{\frac{J_1(\gamma r)}{J_0(\gamma r)}\right\} - \ln\{-\ln(J_0(\gamma r))\}\right] \tag{3.20}$$

Using the above Eq. (3.20), the radial potential profile has been plotted along with the experimental data for different magnetic fields. As seen in the figure, an excellent matching is found between the plasma potential obtained in the experiment and the phenomenological model given by Eq. (3.20).

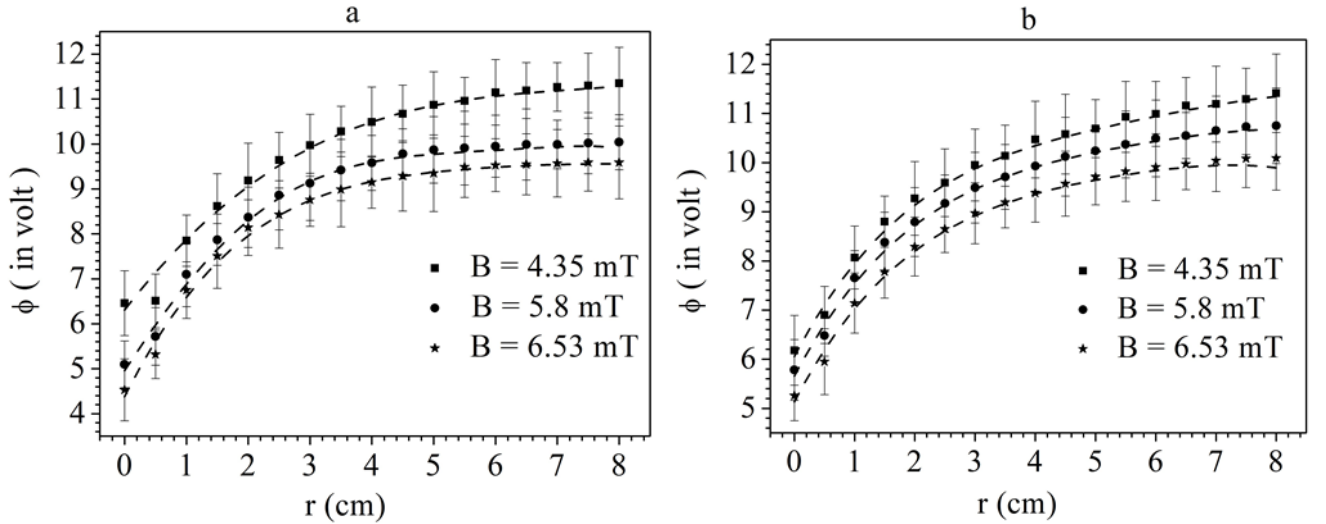

Fig.5. Plot of radial potential variation with different axial magnetic fields using Simon short-circuit effect for P =1.2Pa (a) and 1.7Pa (b). Dash lines are theoretical plots and scattered points are experimental plots.

In Eq. (3.20), the electron temperature $T_e$ has been assumed constant. This fact is substantiated from the experimental data plotted in Fig. 6. $T_e$ is found to remain almost constant over the entire region and it is also unaffected by magnetic field. However its magnitude drops by almost 40 % with increase in pressure from 1.2 to 1.7 Pa. The fall in $T_e$ is likely due to increase in electron-neutral collisions.

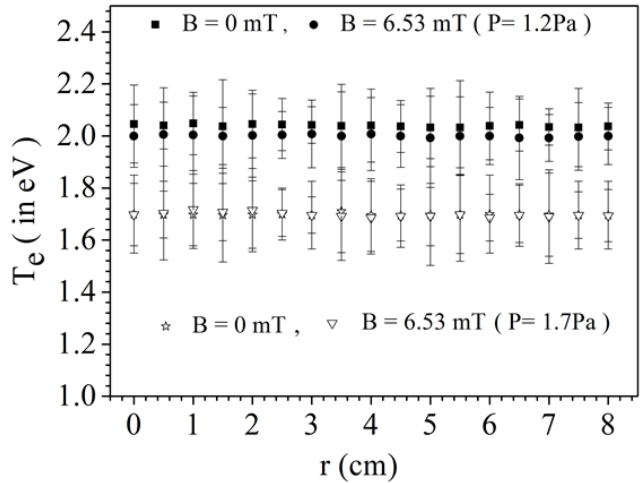

Fig.6. Plot of experimental radial electron temperature variation with and without magnetic field for P = 1.2Pa and 1.7Pa.

## 4. Discussion

The experimental results presented in section-3 clearly demonstrate the effect of magnetic field on the radial distribution of density and plasma potential inside the cylindrical volume. The radial density profile is peaked in the centre, where as a contrasting trend in the radial potential profile showing a marked separation from the normal Boltzmann distribution [c.f figures-4]. It is also observed that with the application of magnetic field, the ∅(r) clearly changes from a Boltzmann to non-Boltzmann behaviour in accordance with the previous report [14].

In fig.7, the radial plasma density and the potential profile for a typical experimental condition are plotted. The contrasting trend observed in the present experiment has also been reported based on hydrodynamic model by previous authors [14]. However the potential profile was seen to increase steeply towards the walls. A similar result was also reported by Fruchtman [28], where they considered the electron distribution to be Boltzmann across the magnetic field. In the present experiment, the plasma potential increases but tend to reach a saturation value with the radial distance from the centre [c.f. fig-5]. This behaviour has been accurately captured through the phenomenological model, which implicitly takes account of short-circuiting effect [16] due to the grounded end plates.

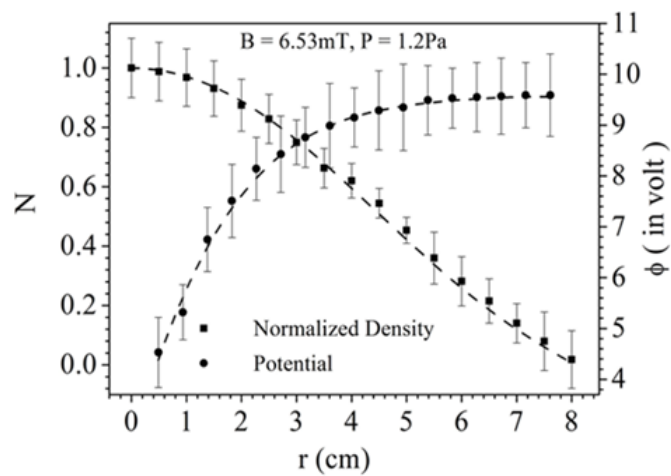

Fig.7. Radial potential and normalized density variation of magnetized plasma for P = 1.2Pa. Dash lines are theoretical plots and scattered points are experimental plots.

In order to verify the significance of the anomalous Bohm diffusion, which assert that the cross-field diffusion across the magnetic field via $\vec{E} \times \vec{B}$ fluctuations, amounts to a diffusion rate $D_{\perp B}$, which is given by;

$$D_\perp = \alpha \frac{k_B T_e}{eB} = D_{\perp e} \qquad (4.1)$$

The empirical factor $\alpha = 1/16$ (~ 0.0625) as prescribed by Bohm [13] has been widely applied, although some authors have also reported higher values of $\alpha$ in the range of 0.21 – 0.4 [30]. If the above diffusion constant is applied to estimate the radial electron flux, it gives us;

$$\Gamma_{\perp e} = -\alpha \frac{k_B T_e}{eB} (\nabla n)_{\perp e} \qquad (4.2)$$

If the short-circuiting effect is ignored, then in order to maintain quasi-neutrality the plasma electrons and positive ions across the magnetic field should leave at the same rate. This compels one to apply ambipolar flow across the magnetic field. Hence equating the perpendicular flux of electrons $\Gamma_{\perp e}$ to the flux of positive ions $\Gamma_{\perp ion} = \frac{1}{4} n_0 v_{\perp ion}$; an expression for $\alpha$ can found as follows;

$$\alpha = -\frac{B\, n_0 v_{\perp ion}}{4 T_e\, (in\ eV)(\nabla n)_{\perp e}} \qquad (4.3)$$

Hence the $\alpha$ can be determined by substituting the quantities determined from the experiment namely; $(\nabla n)_{\perp e}$, $v_{\perp ion}$, $T_e$, $n_0$, $B$ in the equation for a range of pressure and magnetic fields as displayed in Table.1. As seen in the Table-1, the value of $\alpha$ is close to 1/16 as predicted by Bohm.

Table.1

| Pressure (in Pa) | B = 4.35mT | B = 5.8mT | B = 6.53mT |
| --- | --- | --- | --- |
| 1.2 | $\alpha = 0.064$ | $\alpha = 0.060$ | $\alpha = 0.063$ |
| 1.7 | $\alpha = 0.069$ | $\alpha = 0.067$ | $\alpha = 0.062$ |

The experimental conditions are able to predict the anomalous Bohm diffusion; however at the same time the radial potential profile is accurately determined from the phenomenological model via Eq. (3.16), which takes in to the short circuiting effect by the grounded end plate. Therefore the above observation points towards a possible links between the two underlying mechanisms.

While in the case of short-circuiting effect, there is no reliance on ambipolar flow assumption across the magnetic field; however the effect can virtually increase the perpendicular electron flux $\Gamma_{\perp e}$ across the magnetic field, making it apparently ambipolar. While arriving at Eq. 3.20, we have considered the total flux balance where the parallel flux of electrons, obeying Boltzmann distribution comes in the picture. This could be a possible reason why the ambipolarity assumption and Boltzmann distribution across magnetic field lines has also yielded valid results [19, 24, 25]. While the short-circuit effect can enhance the radial electron flux by allowing a net current flowing through the conducting end-plate. This is possible to achieve by having a spread of radial potential Ø(r) above the grounded end plate. Hence the thermal flux of electrons arriving from the different regions inside the magnetic flux-tube will also have spread at different radial positions on the plate. As found from fig-7, the plasma potential increases with the radial distance; thus the electron flux experience a more retarding field at radial grounded endplate as compared with the centre. Consequently, the electron loss along the axis is rapidly increases at the centre whereas it reduces with radial distance. The excess electronic charge at the centre is compensated by the positive ions absorbed by the wall, setting up a current through the end-plate. It is also true that the short-circuit effect alone may not be entirely enough to enhance the electron flow across the magnetic field; but also requires a mechanism to impede the positive ions flowing across the magnetic field lines. Fortunately this is self-consistently achieved as a radial potential well is created at the centre.

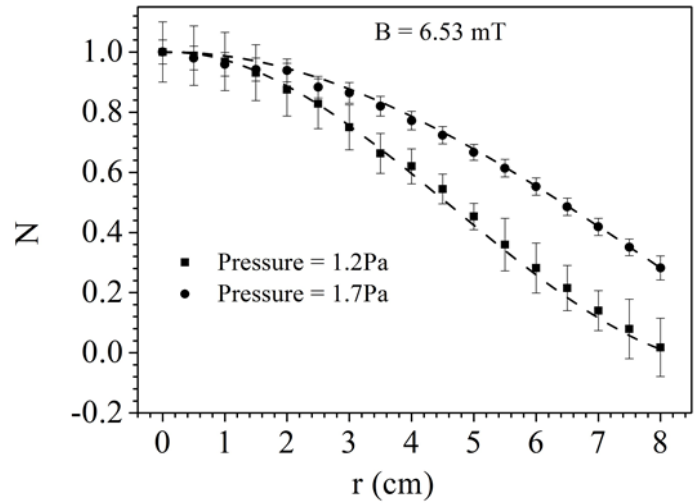

Fig.8. Radial normalized density variation with different gas pressures. Dash lines are theoretical plots and scattered points are experimental plots.

It is also observed in Fig. 8, that the radial plasma density tends to become more flat as the pressure increases. This is caused due to enhancement in cross-field diffusion of electrons across the magnetic field lines.

## 5. Conclusion

Summarizing the overall contents of the paper, a phenomenological model has been formulated to explain the radial plasma density and the potential profile due to axial magnetic field inside a cylindrical plasma column. It has been shown that the Boltzmann treatment is no longer valid across the magnetic field, but due to the short-circuiting effect the radial potential is indirectly linked with the parallel electron flux which obeys Boltzmann distribution. Though the $\vec{E} \times \vec{B}$ induced charge particle fluctuation which leads to 1/B scaling of the diffusion rate; however the excellent matching of the experimental results with the phenomenological model allows us to arrive at a conclusion that at least in the present experimental setup, the Simon short-circuit effect is mainly responsible for observing the contrasting behaviour in the density and the plasma potential.

The advantage of using axial magnetic field in linear plasma devices is that by confining the electrons at the centre of the plasma column one can also confine the ions electro-statically at the centre of the discharge. It is also important that in many discharge setups magnetic field is applied to enhance the plasma density; however it can result in undesirable effects such as in-homogeneity in ion density or ion energy across the substrate [2]. In fusion plasma devices, the role of the conducting electrode on the plasma behaviour near the diverter, limiter or RF antenna is also important. By influencing the radial electric field, one can enhance or suppress various $\vec{E} \times \vec{B}$ instabilities arising inside the plasma. The analysis and the experimental results presented in this paper may elucidate some of the possible effects while dealing such systems.

**Acknowledgements:**

The author thanks Dr. Joydeep Ghosh for his valuable suggestions regarding the manuscript and my colleagues for their supports to do the experiment.